# Field-Angle Dependence of Phonon Thermal Hall Effect in Na$_2$X$_2$TeO$_6$ (X = Co, Zn)

Jian Yan[1,2], Hikaru Takeda[2], Haruka Iwahata[2], Jun-ichi Yamaura[2], Rajesh Kumar Ulaganathan[3,4], Kalaivanan Raju[3,5], Raman Sankar[3], and Minoru Yamashita[2]

1 Institute for Advanced Study, Chengdu University, Chengdu, Sichuan, 610106, China
2 The Institute for Solid State Physics, The University of Tokyo, Kashiwa, 277–8581, Japan
3 Institute of Physics, Academia Sinica, Taipei 11529, Taiwan
4 Centre for Nanotechnology, Indian Institute of Technology, Roorkee, 247667, India
5 Department of physics and chemistry, Prairie View A&M University, Prairie View, TX 77446, USA

Corresponding Authors: Jian Yan (yanjian@cdu.edu.cn) and Minoru Yamashita (my@issp.u-tokyo.ac.jp)

**Abstract:**

The mechanism behind thermal Hall effects by phonons, which are observed in various materials, is not clarified despite the dominant contribution as heat carriers. Theoretically, mechanisms based on the intrinsic Berry phase and those on extrinsic impurity-induced scatterings have been proposed, which can be distinguished by comparing the field-angle dependence of the thermal Hall effect and that of the magnetic anisotropy. Here, we investigate the field-angle dependence of the thermal Hall effects in the antiferromagnet Na$_2$Co$_2$TeO$_6$ and its non-magnetic isostructural analogue Na$_2$Zn$_2$TeO$_6$ in the $ac$ plane. We find that the field-angle dependence of the thermal Hall conductivity in both materials well follows that of the out-of-plane magnetization, showing a common mechanism by extrinsic impurity-induced scatterings in both the phonon thermal Hall effect and that enhanced by a coupling with the magnetism.

# INTRODUCTION:

Thermal Hall effects (THEs) in an insulator, which should be forbidden owing to the apparent absence of conduction electrons, have been observed in various insulators, including ferromagnets [1–3], antiferromagnets [4–6], Kitaev candidate materials [7–12], spin ice [13], high-$T_c$ cuprates [14–16] and even in non-magnetic insulators [17–21]. As the origin behind these THEs, not only magnetic excitations, such as magnons [1–3,6,11], spinons [4,5], and Majorana fermions [7–9] but also non-magnetic phonons [5,10,12–22] are suggested. Among these possible origins, given the pervasive presence of the phonon thermal conduction in all materials, it is especially important to clarify the underlying mechanism of the phonon THEs to discriminate other THEs by magnetic excitations.

However, the understanding of phonon THEs is far behind those for magnetic excitations. This is because, whereas THEs exhibited by these magnetic excitations are basically understood in terms of the Berry curvature of their energy bands, it is not even clear how the dynamics of phonons are coupled with the magnetic field. For example, the magnon thermal Hall conductivity ($\kappa_{xy}$) observed in the magnetic skyrmion phase can be well described by the Berry curvature of the magnon bands given by the magnetic skyrmion lattice [3,6]. The half-quantized $\kappa_{xy}$ in the Kitaev model is also realized by the Berry curvature of the Majorana fermions, which is suggested to be observed in α-RuCl₃ [7–9]. In contrast, for the intrinsic Berry curvature [23,24] and extrinsic impurity-induced scatterings [25–32] mechanisms suggested for phonon THEs, the effects of the intrinsic Berry curvature are generally too weak to account for the magnitude of $\kappa_{xy}$ measured in various materials, and it is not clear if the calculations based on the extrinsic mechanism are consistent with the common thermal Hall angle ($\kappa_{xy}/\kappa_{xx}$) observed in various materials [19], requiring a new angle of study to clarify the phonon THE.

Here, we suggest that the magnetic field-angle dependence of $\kappa_{xy}$ plays an important role in distinguishing between the intrinsic and extrinsic mechanisms of the phonon THE. The field-angle dependence of $\kappa_{xy}$ from an intrinsic mechanism is given by the magnetic anisotropy of the spin Hamiltonian of the system, as the field-angle dependence of $\kappa_{xy}$ in the Kitaev model is determined by the magnetic field direction with respect to the spin axes [9]. On the other hand, the field-angle dependence of $\kappa_{xy}$ from an extrinsic mechanism is given by the angle between the heat flow and the magnetization (**M**) as that in the anomalous Hall effect in ferromagnetic metals [33]. For example, in the theoretical studies of extrinsic skew scatterings [25,28], the field-angle dependence of $\kappa_{xy}$ is suggested to be given by the scattering rate $W_{\mathbf{k}\to\mathbf{q}} \propto \mathbf{M} \cdot (\mathbf{k} \times \mathbf{k}')$, where $\mathbf{k}$ and $\mathbf{k}'$ are the phonon momentum of the initial and the final state, respectively. In this case, when the heat current is applied in the $x$–$y$ plane, the field-angle dependence of $\kappa_{xy}$ becomes proportional to that of the out-of-plane magnetization, i.e. $\kappa_{xy}(\theta) \propto M_c(\theta)$, where $\theta$ denotes the angle of the magnetic field from the axis perpendicular to the basal plane and $M_c$ is the magnetization along the $c$ ($\parallel z$) axis (see Fig. 1(b)). Therefore, the field-angle dependence of a phonon THE will provide new information to facilitate the understanding for the underlying mechanism. In addition, one needs to consider the possibility that phonon THEs may have multiple mechanisms as in the anomalous Hall effect in ferromagnetic metals [33]. Therefore, the phonon THEs in different materials may be caused by different mechanisms depending on the details of the material structure, requiring comparative studies done in the isostructural compounds.

In this study, we investigate the magnetic field-angle dependence of $\kappa_{xy}$ and the longitudinal thermal conductivity ($\kappa_{xx}$) in the Kitaev-candidate antiferromagnet $Na_2Co_2TeO_6$ (NCTO) and its isostructural non-magnetic analogue $Na_2Zn_2TeO_6$ (NZTO). Our $\kappa_{xy}$ measurements done on both magnetic and non-magnetic isostructural insulators allow us to discriminate the phononic and magnetic contributions in $\kappa_{xy}$. In the non-magnetic NZTO, we reveal that $\kappa_{xy}$ by phonons shows a clear field-angle dependence of $\kappa_{xy}(\theta) \propto \cos\theta$, suggesting a dominant contribution by extrinsic impurity-induced scatterings. In the paramagnetic phase of the magnetic NCTO, the phonon THE is observed to be enhanced by a coupling with the magnetism. We find that $\kappa_{xy}(\theta)$ of this enhanced phonon THE also follows $M_c(\theta)$ by considering its easy-plane anisotropy, which is consistent with the extrinsic effects. We suggest that the phonon THE in both compounds is commonly caused by extrinsic skew scatterings, on the basis of the similarity to the anomalous Hall effect in ferromagnetic metals.

**RESULTS:**

The cobalt-based antiferromagnet NCTO constitutes the two-dimensional honeycomb layers of $Co^{2+}$ ions (Fig. 1a) which form $J_{eff} = 1/2$ Kramers doublets [34–41]. The temperature dependence of the magnetic susceptibility ($\chi(T)$) shows a large easy-plane anisotropy ($\chi_a > \chi_c$) and an anomaly by the antiferromagnetic order at the Néel temperature $T_N = 27$ K (Fig. 1d). This antiferromagnetic order can be suppressed by applying magnetic field parallel to the $ab$ plane, which results in the increase of $\kappa_{xx}$ at 15 T applied parallel to the $a$ axis (Fig. 1c) [11,42]. In the isostructural non-magnetic analogue NZTO, the magnetic $Co^{2+}$ ions are replaced by the non-magnetic $Zn^{2+}$ with keeping the lattice structure of NCTO ($P6_322$) including the disorder of Na ions [35,36,43]. The magnitude of $\chi$ of NZTO is more than two orders of magnitude smaller than $\chi$ of NCTO and is dominated by the core diamagnetic contribution for both $B \parallel ab$ and $B \parallel c$ (see Fig. S1 in Supplementary Material (SM)), confirming that NZTO is non-magnetic. The temperature dependence of $\kappa_{xx}$ of NZTO shows the phonon peak around 25 K (Fig. 1c) with a very small field dependence as shown below (Fig. 2).

We first investigate the field dependence of the normalized magnetothermal conductivity $\Delta\kappa_{xx}(B)/\kappa_{xx}(0) = (\kappa_{xx}(B) - \kappa_{xx}(0))/\kappa_{xx}(0)$ and that of the thermal Hall conductivity divided by the temperature $\kappa_{xy}/T$ in NZTO in which only the phonons are heat carriers. Figure 2 shows the representative data at various temperatures and angles. As shown in Figs. 2(a–c), the field dependence of $\Delta\kappa_{xx}(B)/\kappa_{xx}(0)$ at 20 K shows a positive magnetothermal conduction at 0° that becomes smaller at higher angles. This positive magnetothermal conduction becomes further smaller at higher temperatures, which is followed by the negative magnetothermal conduction for all the angles at 45 K. This positive magnetothermal conduction is typical of the phonon thermal conduction that is enhanced by suppressing the scattering by magnetic impurities under magnetic field. The negative magnetothermal effect observed at higher temperatures would be due to a resonant scattering effect between the phonons and the Zeeman gap, which is known to block the phonon conduction [44]. Therefore, despite the non-magnetic property of NZTO, the field dependence of $\Delta\kappa_{xx}(B)/\kappa_{xx}(0)$ shows the presence of weak scattering effects on the phonons due to residual magnetic impurities in the sample.

In contrast to the field dependence of $\Delta\kappa_{xx}(B)/\kappa_{xx}(0)$, only negative $\kappa_{xy}/T$ is observed for all the fields and angles with a linear field dependence (Figs. 2(d–f)), as observed in other non-magnetic materials [18,19]. As described later, the magnitude of $\kappa_{xy}/T$ decreases with increasing the temperature (Fig. 4(b)) and the field angle from $B \parallel c$ to $B \parallel ab$ (Fig. 5).

Next, we examine the field dependence of $\Delta\kappa_{xx}(B)/\kappa_{xx}(0)$ and that of $\kappa_{xy}/T$ in NCTO above 20 K (see Fig. S2 in SM for the lower temperature data). As shown in Fig. 3, both the magnitude of $\Delta\kappa_{xx}(B)/\kappa_{xx}(0)$ and that of $\kappa_{xy}/T$ are substantially larger than those in NZTO, showing enhancement effects by the magnetic component. As shown in Figs. 3(a) and 3(b), $\Delta\kappa_{xx}(B)/\kappa_{xx}(0)$ shows the positive magnetothermal conductivity for the in-plane field, which turns into the negative one for the out-of-plane field. We note that the increase of $\Delta\kappa_{xx}(B)/\kappa_{xx}(0)$ at 22 K under the in-plane field matches well with the results of the earlier work done in higher fields [11] (see Fig. S3 in SM). At higher temperatures, only the negative magnetothermal conductivity is observed for all the angles as shown in Fig. 3(c). Although the positive $\Delta\kappa_{xx}(B)/\kappa_{xx}(0)$ can be understood by the same field-enhancing effect on phonons observed in NZTO, the negative one may contain a decrease of a magnetic contribution in $\kappa_{xx}$ as suggested by the previous work [10] (see SM for details).

The dependence of $\kappa_{xy}/T$ on the field and the angle in NCTO is also quite different from that of $\Delta\kappa_{xx}(B)/\kappa_{xx}(0)$. At 22 K and at 0°, the negative $\kappa_{xy}/T$ starts to increase above around 4 T (Fig. 3(d)) in contrast to the immediate field-induced decrease of $\Delta\kappa_{xx}(B)/\kappa_{xx}(0)$ (Fig. 3(a)). This negative $\kappa_{xy}/T$ is enhanced for 45°–60°, which is followed by a rapid decrease at 90° (Fig. 3(d)). At 30 K, on the other hand, $\kappa_{xy}/T$ shows a linear increase to the magnetic field up to ~8 T except $\theta = 90°$ (Fig. 3(e)) despite the absence in the field dependence of $\Delta\kappa_{xx}(B)/\kappa_{xx}(0)$ below 6 T for 0°– 30° (Fig. 3(b)). The field-angle dependence of $\kappa_{xy}/T$ is not discernible at 30 K for 0°–45°, which is followed by a gradual decrease as approaching $\theta = 90°$ (Fig. 3(e)). The dependence of $\kappa_{xy}/T$ on the field and the angle at 46 K (Fig. 3(f)) is similar to that at 30 K, in contrast to the disappearance of the positive magnetothermal conduction at this temperature (Fig. 3(c)). This contrasting field and field-angle dependence of $\Delta\kappa_{xx}(B)/\kappa_{xx}(0)$ and that of $\kappa_{xy}/T$ indicate that the field-induced changes in the scattering on the heat carriers are irrelevant to the THEs in NCTO. The different field dependence of $\kappa_{xy}/T$ at 22 K ($< T_N$) from that at 30 and 46 K is attributed to an effect by the AFM order as discussed in SM (see Fig. S7 in SM).

We summarize the temperature dependence $-\kappa_{xy}/TB$ and $\kappa_{xx}/T$ of NCTO and NZTO in Fig. 4(a) and Fig. 4(b), respectively. We also plot the field-angle dependence of $-\kappa_{xy}/TB$ and $\Delta\kappa_{xx}(B)/\kappa_{xx}(0)$ extracted from Fig. 2 and Fig. 3 at a constant temperature in Fig. 5 (30 K), Fig. S7 (20 K), and Fig. S8 (45 K).

## DISCUSSION:

We first discuss the enhancement of the phonon THE in the paramagnetic phase of NCTO by the magnetic component. As shown in Fig. 4(b), the temperature dependence of $-\kappa_{xy}/TB$ of NZTO at $\theta = 0$ depends almost linearly on that of $\kappa_{xx}/T$, which is characteristic of a phonon THE as observed in other non-magnetic insulators [18,19]. Remarkably, this linear scaling is also observed in the magnetic NCTO as shown in Fig. 4(a), suggesting a dominant phonon contribution in $\kappa_{xy}$ of NCTO. In addition, $|\kappa_{xy}/TB|$ of NCTO is larger than that of the isostructural NZTO for all angles, showing an enhancement of the phonon $\kappa_{xy}$ by the magnetism in NCTO. The presence of the magnetic contribution in $\kappa_{xy}$ is also supported by the negative magnetothermal conduction (Figs. 3(a)–(c)). These results suggest that $\kappa_{xy}$ of NCTO comes from hybrids of phonons and magnetic excitations [45–47]. A similar enhancement of $\kappa_{xy}$ is also reported in the two-dimensional van der Waals magnet VI$_3$ [2].

Next, we discuss the field-angle dependence of the THEs observed in NCTO and NZTO. As shown in Fig. 5, the field-angle dependence of the phonon $\kappa_{xy}$ in NZTO fits well to the cosine function (the grey dashed line). We note that the slight deviation of the data from the cosine function would be due to the scatters of the data (see Figs. 2(d)–(f) and Fig. S6 in SM). On the other hand, the field-angle dependence of $\Delta\kappa_{xx}(B)/\kappa_{xx}(0)$ stays almost constant in the whole field-angle range, showing negligible field-angle dependence of the phonon scattering intensity. Therefore, we conclude that the cosine field-angle dependence of $\kappa_{xy}$ in NZTO is not consistent with an intrinsic mechanism because there should be no field-angle dependence in an intrinsic mechanism due to the isotropic non-magnetic property of NZTO (Fig. 1(d)). On the other hand, the cosine field-angle dependence of $\kappa_{xy}$ is consistent with an extrinsic impurity-induced mechanism that the field-angle dependence of the phonon THE is determined by the angle between the heat flow and the magnetization of the residual magnetic impurities that are observed to affect $\kappa_{xx}$ (Figs. 2(a–c)).

The field-angle dependence of $-\kappa_{xy}/TB$ of NCTO shows a sharper decrease at angles greater than 60°, reaching to zero at 90°. This angle dependence is more pronounced than the cosine function of $-\kappa_{xy}/TB$ of NZTO. Meanwhile, the field-angle dependence of $\Delta\kappa_{xx}(B)/\kappa_{xx}(0)$ of NCTO is as small as that of NZTO. This field-angle dependence of $-\kappa_{xy}/TB$ in NCTO can also be understood by the same extrinsic mechanism in NZTO. Indeed, the field-angle dependence of $|\kappa_{xy}/TB|$ of NCTO fits well with an elliptic angle dependence of the out-of-plane magnetization $M_c$ (the red dashed line) by considering the easy-plane anisotropy such that $\kappa_{xy}(\theta) \propto M_c(\theta)$, where $M_c(\theta) = M_c(0) \cdot (1 + (\tan\theta/\alpha)^2)^{-\frac{1}{2}}$ and $\alpha = 3.2$ is the averaged ratio of the magnetization along the $a$ axis to that along the $c$ axis (Fig. 1(d)), indicating that the phonon THE in the paramagnetic phase of NCTO also comes from extrinsic scatterings given by the magnetization perpendicular to the basal plane. We also find a similar scaling between the angle dependence of $|\kappa_{xy}/TB|$ and that of the out-of-plane magnetization at 20 K (Fig. S7) and 45 K (Fig. S8) for both NCTO and NZTO, except for the data in the antiferromagnetic phase of NCTO at 20 K for $\theta < 45°$, where a suppression of $|\kappa_{xy}/TB|$ from the elliptic angle dependence is observed due to an emergence of another magnetic contribution in the antiferromagnetic phase. Therefore, we conclude that the phonon THE observed in both NCTO (the paramagnetic phase) and NZTO has its origin in the extrinsic phonon-magnetism scatterings.

In the anomalous Hall effects in ferromagnetic metals [33], the anomalous Hall conductivity by the extrinsic skew scatterings is shown to have the field-angle dependence given by the scattering rate $W_{\mathbf{k} \to \mathbf{q}} \propto \mathbf{M} \cdot (\mathbf{k} \times \mathbf{k}')$, which is consistent with the field-angle dependence of the phonon $\kappa_{xy}$ observed in our measurements. From this analogy, we suggest that similar skew scatterings of phonons by the magnetism play an important role in the phonon THE. In fact, the theoretical works in Ref. [25] and Ref. [28] point out the importance of a similar skew scattering given by the angle between the magnetization and the thermal current. We note, however, that the situations considered in these works differ from that studied in our materials; the former considers the effects by the quadrupole moment of the superstoichiometric $Tb^{3+}$ ions and the latter the phonon scatterings by charge defects, requiring further theoretical studies to identify the details of the extrinsic mechanism behind the linear correlation between $\kappa_{xy}(\theta)$ and $M_c(\theta)$.

Moreover, similar to the three mechanisms (intrinsic, extrinsic skew, and extrinsic side-jump) considered in the anomalous Hall effects, there might be multiple mechanisms behind the phonon THE, and one of them may emerge as a dominant mechanism depending on the material conditions. In fact, a large sample dependence in the planar $\kappa_{xy}$ at $\theta = 90°$ is reported in different NCTO samples [12], implying a presence of another mechanism for the planar $\kappa_{xy}$ related to the sample quality, in addition to the phonon-magnetism scattering by $M_c(\theta)$ dominating the field-angle dependence of $\kappa_{xy}(\theta)$ found in this work. Moreover, the phonon THE by resonance scatterings has been reported in the metallic spin ice compound [13], which is suggested to be explained by a side-jump mechanism by non-Kramers ions [31]. The different field-angle dependence reported in the non-magnetic black phosphorus [19,21] may also be caused by a different mechanism acting for the ballistic phonons in the highly clean sample. Therefore, as a key future challenge, it is important to investigate other mechanisms by examining an intrinsic contribution through first-principles calculations or incorporating inelastic X-ray scattering to directly probe topological nature of phonon bands, as well as by extending the study of field-angle dependence of thermal Hall effects to other materials with strong spin-orbit couplings (e.g. α-RuCl$_3$).

In summary, we investigate the field-angle dependence of the thermal conductivity and the thermal Hall conductivity in the $ac$ plane for both the antiferromagnet Na$_2$Co$_2$TeO$_6$ and its isostructural non-magnetic analogue Na$_2$Zn$_2$TeO$_6$. Our measurements done in isostructural magnetic and non-magnetic compounds allow us to study the purely phonon contribution in $\kappa_{xy}$ and that coupled with the magnetism. We reveal that both the field-angle dependence of $\kappa_{xy}$ of the paramagnetic phase in Na$_2$Co$_2$TeO$_6$ and that in Na$_2$Zn$_2$TeO$_6$ follow that of the out-of-plane magnetization, suggesting the presence of a common extrinsic mechanism by phonon-magnetism scatterings. Given the similarity to the anomalous Hall effect in ferromagnetic metals, we suggest that the field-angle dependence of $\kappa_{xy}$ in both compounds is caused by extrinsic skew scatterings.

**Acknowledgement:**

We thank K. Behnia for the fruitful discussions. J.Y. was supported by Grant-in-Aid for JSPS Fellows. J.Y. and M.Y. thank the support by JSPS KAKENHI Grant Numbers JP22KF0111 and JP23H01116. R.S. acknowledges the financial support provided by the Ministry of Science and Technology in Taiwan under Project No. NSTC-113-2124-M-001-003 and No. NSTC-113-2112M001-045-MY3, as well as support from Academia Sinica for the budget of AS-iMATE11412. R.K.U. would like to acknowledge the IITR for the Faculty Initiation Grant (FIG-101068).

**Author contribution:**

Jian Y. and M.Y. conceived the project. Jian Y. and H.T. performed thermal transport measurements. R.K.U., K.R., and R.S. synthesized and characterized the samples. Jun-ichi Y. performed X-ray measurements. Jian Y., H.T., H.I. and R.S. performed magnetic susceptibility measurements. M.Y. wrote the paper in consultation with all coauthors.

## Data availability:

All the data supporting this study is available from the corresponding authors upon reasonable request.

## Competing interests:

Authors declare that they have no competing interests.

**FIGURES and LEGENDS:**

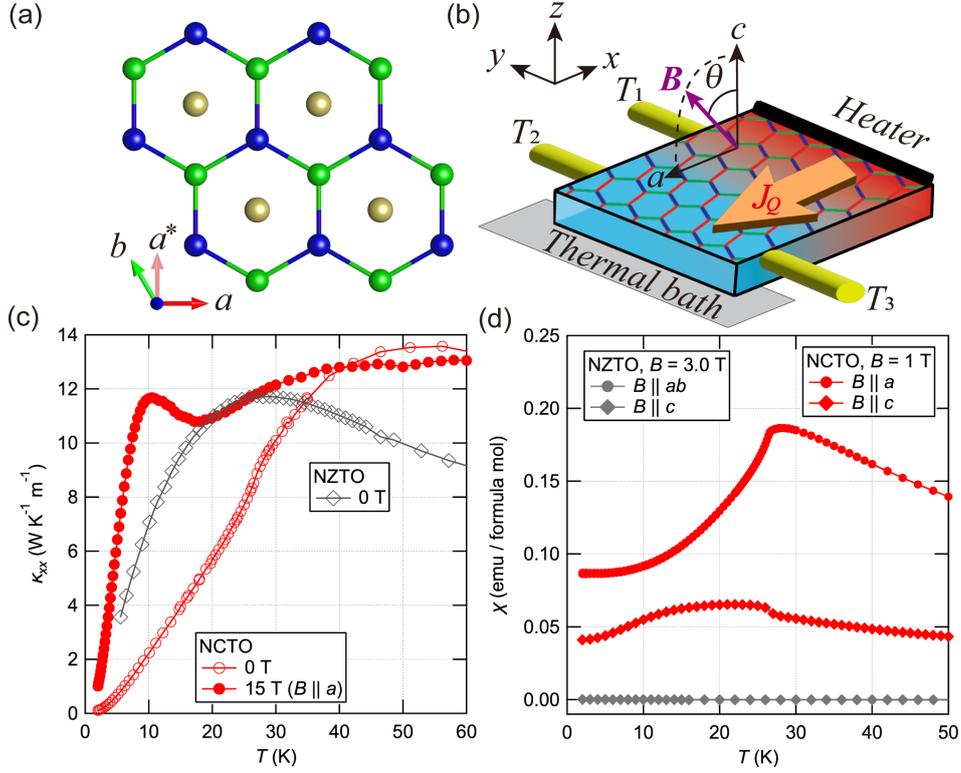

Fig. 1 (a) A schematic of the crystal structure showing Co/Zn (blue and green circles) and Te (yellow) atoms in the $ab$ plane. (b) An illustration of our experimental setup for the thermal conductivity and the thermal Hall measurements. The temperature gradients caused by the heat current $J_Q$ applied along the $a$ axis were measured by the three thermometers. The magnetic field $B$ was applied in the direction of $\theta$ from the $c$ axis in the $ac$ plane (see SM for more details of the setup). (c,d) The temperature dependence of the longitudinal thermal conductivity ($\kappa_{xx}$, c) and the magnetic susceptibility ($\chi$, d) of NCTO (red) and NZTO (grey). The data of $\kappa_{xx}$ and $\chi$ in $B \parallel a$ of NCTO is taken from our previous work [11]. See Fig. S1 in SM for an enlarged view of $\chi$ of NZTO.

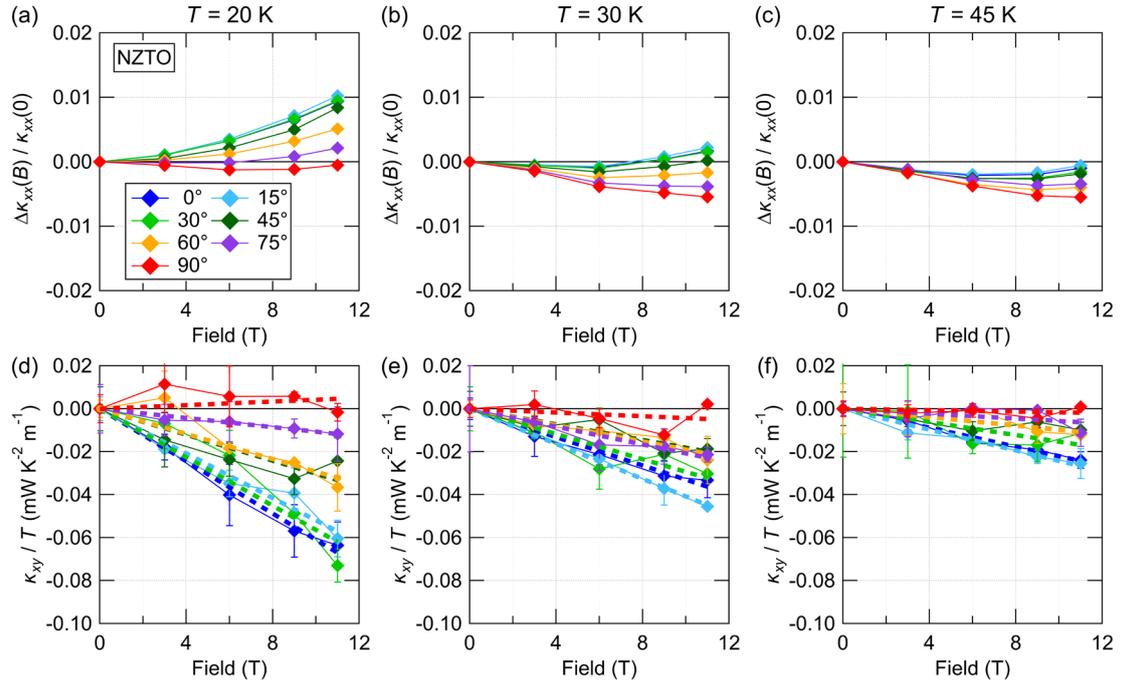

Fig. 2 The magnetic field dependence of $\Delta\kappa_{xx}(B)/\kappa_{xx}(0)$ (a–c) and $\kappa_{xy}/T$ (d–f) at 20 K (a, d), 30 K (b, e), and 45 K (c, f) and at different angles of NZTO. The averaged data measured in the magnetization and the demagnetization processes is plotted. The error bars show the deviation of the data in the two processes. The error bars are not shown for the data of which the errors are smaller than the symbol size as those in $\Delta\kappa_{xx}(B)/\kappa_{xx}(0)$. The dashed lines in (d–f) show a linear fit of the $\kappa_{xy}/T$ data.

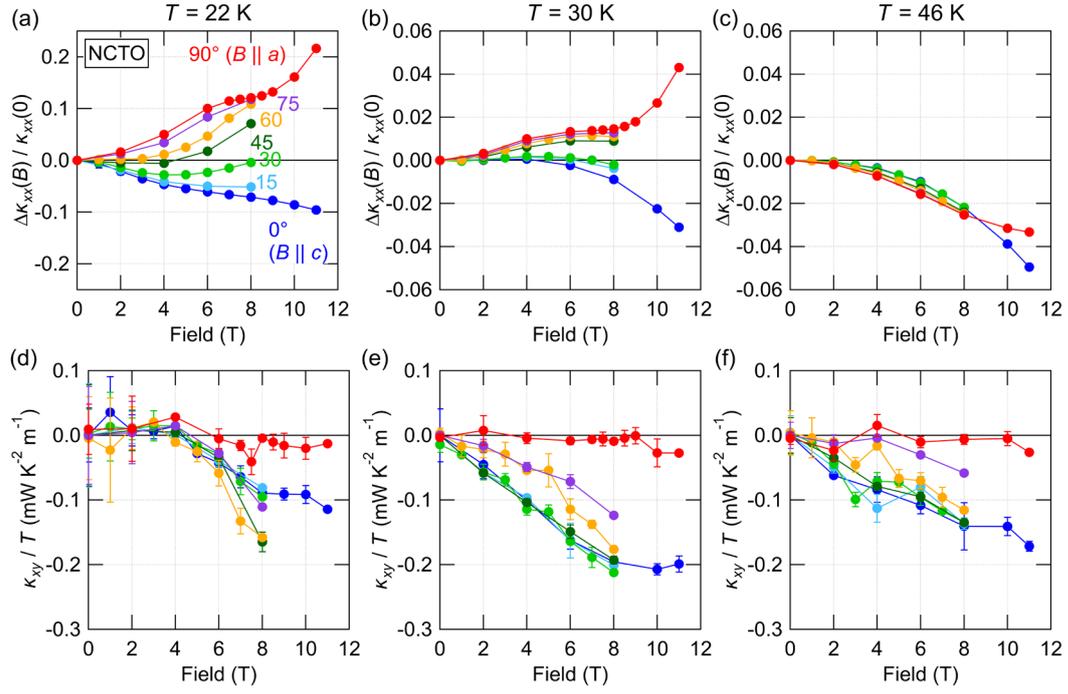

Fig. 3 The magnetic field dependence of $\Delta\kappa_{xx}(B)/\kappa_{xx}(0)$ (a–c) and $\kappa_{xy}/T$ (d–f) at 22 K (a, d), 30 K (b, e), and 46 K (c, f) and at different angles of NCTO. The error bar definition is the same as that used in Fig. 2.

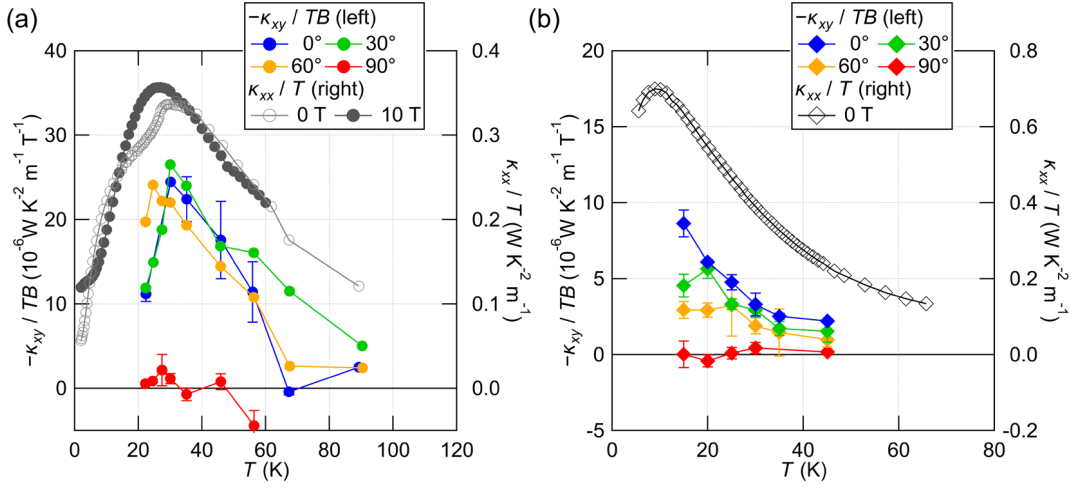

Fig. 4 The temperature dependence of $-\kappa_{xy}/TB$ (left axis) and $\kappa_{xx}/T$ (right axis) of NCTO (a) and NZTO (b). The data at 8 T is used to estimate $-\kappa_{xy}/TB$ of NCTO, whereas the slope of the linear fit (the dashed lines in Fig. 2) is used for NZTO. The error bar definition of the NCTO data is the same deviation as used in Fig. 3. The error bars of the NZTO data show the greater of the same deviation at 9 T or the standard deviation of the linear fit. The data of $\kappa_{xx}/T$ at 10 T applied along the $a$ axis is plotted for NCTO to show the temperature dependence of the phonon contribution without the effect by the AFM order that develops below $T_N$. See Figs. S4–S6 in SM for the additional field dependence data of $\kappa_{xy}/T$.

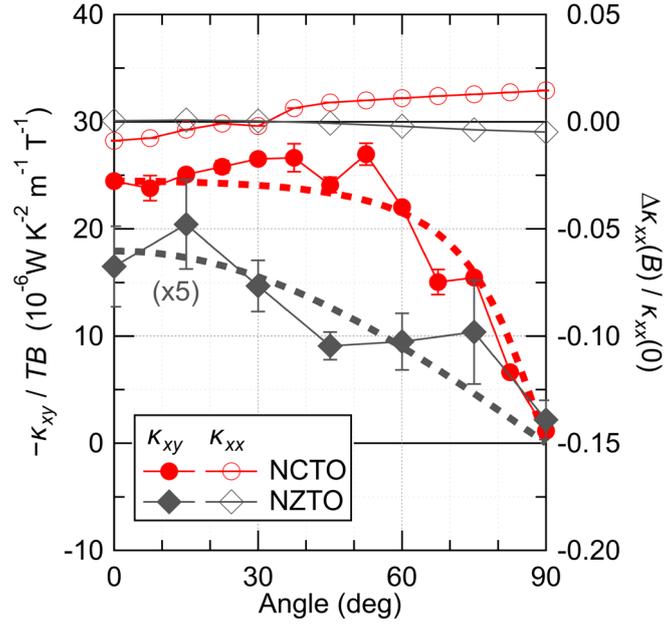

Fig. 5 The field-angle dependence of $-\kappa_{xy}/TB$ (filled symbols, left) and $\Delta\kappa_{xx}(B)/\kappa_{xx}(0)$ (open symbols, right) of NCTO (red) and NZTO (grey) at 30 K. The $\Delta\kappa_{xx}(B)/\kappa_{xx}(0)$ data at 8 T (9 T) is shown for NCTO (NZTO). The $-\kappa_{xy}/TB$ data of NZTO is multiplied by 5 for clarity. The error bar definitions are the same as those used in Fig. 4. The grey dashed line shows a cosine fit to the $-\kappa_{xy}/TB$ data of NZTO. The red dashed line shows a fit to the $-\kappa_{xy}/TB$ data of NCTO by the field-angle dependence of the out-of-plane magnetization.

# Supplementary Material for

# Field-Angle Dependence of Phonon Thermal Hall Effect in

# Na$_2$X$_2$TeO$_6$ (X = Co, Zn)

Jian Yan, Hikaru Takeda, Haruka Iwahata, Jun-ichi Yamaura, Rajesh Kumar Ulaganathan, Kalaivanan Raju, Raman Sankar, and Minoru Yamashita

This Supplementary Material includes
- Materials and Methods
- Fig. S1, The temperature dependence of the magnetic susceptibility of Na$_2$Zn$_2$TeO$_6$
- The field dependence of the magnetothermal conductivity in Na$_2$Co$_2$TeO$_6$ (NCTO) and Na$_2$Zn$_2$TeO$_6$ (NZTO)
- Figs. S2–S6, the additional field dependence data of $\kappa_{xx}$ and $\kappa_{xy}$ of Na$_2$Co$_2$TeO$_6$ and Na$_2$Zn$_2$TeO$_6$
- Figs. S7 and S8, the additional field-angle dependence of $-\kappa_{xy}/TB$ and $\Delta\kappa_{xx}(B)/\kappa_{xx}(0)$ of Na$_2$Co$_2$TeO$_6$ and Na$_2$Zn$_2$TeO$_6$ at around 20 K and at 45 K

**Materials and Methods**

The crystals of both Na$_2$Co$_2$TeO$_6$ and Na$_2$Zn$_2$TeO$_6$ used in this study were synthesized by the self-flux method as described in Ref. [1]. For the thermal-transport measurements of Na$_2$Co$_2$TeO$_6$, the same single crystal used in Ref. [2] was used.

The thermal-transport measurements were performed by attaching one heater and three thermometers to the sample with a silver paste or a stycast (2850FT) as shown in Fig. 1(b) in the main text. To avoid a background signal coming from a metal, the sample was attached to the insulating LiF heat bath with non-metallic grease. The temperature difference along (perpendicular) to the heat current $J_Q = Q/wt$ applied along the $a$ axis was measured by the three thermometers as $\Delta T_x = T_1 - T_2$ ($\Delta T_y = T_2 - T_3$), where $Q$ is the heater power, $w$ is the sample width, and $t$ is the sample thickness. The sample temperature $T$ was determined by the average of $T_1$ and $T_2$, such that $T = (T_1 + T_2)/2$. The magnetic field $B$ was applied in the direction of $\theta$ from the $c$ axis in the $ac$ plane. The field-angle dependence of $\kappa_{xx}$ and $\kappa_{xy}$ was measured by rotating the measurement cell in the magnetic field. To cancel the longitudinal component in $\Delta T_y$ by the misalignment effect, $\Delta T_y$ was antisymmetrized with respect to the field direction as $\Delta T_y^{asym} = (\Delta T_y(+B) - \Delta T_y(-B))/2$. To take into account a magnetic hysteresis effect, this antisymmetrization was done separately for the data obtained in the field-up process and that in the field-down process. The thermal (Hall) conductivity $\kappa_{xx}$ ($\kappa_{xy}$) is derived by

$$\begin{pmatrix} Q/wt \\ 0 \end{pmatrix} = \begin{pmatrix} \kappa_{xx} & \kappa_{xy} \\ -\kappa_{xy} & \kappa_{xx} \end{pmatrix} \begin{pmatrix} \Delta T_x/L \\ \Delta T_y^{asym}/w \end{pmatrix},$$

where $L$ is the length between $T_H$ and $T_{L1}$. We note that, whereas the two-fold rotation axis symmetry around the $a^*$ axis prohibits a finite $\kappa_{xy}$ for $B \parallel a^*$ in both NCTO and NZTO [2], it is not our case for $B \parallel a$.

The magnetic susceptibility was measured by using a superconducting quantum interference device (SQUID) magnetometer (Quantum Design, magnetic property measurement system (MPMS)).

**The temperature dependence of the magnetic susceptibility of Na$_2$Zn$_2$TeO$_6$ (NZTO)**

Figure S1 shows an enlarged view of the magnetic susceptibility ($\chi$) data of NZTO. As shown in Fig. S1, the magnitude of $\chi$ is more than two orders of magnitude smaller than $\chi$ of Na$_2$Co$_2$TeO$_6$ (NCTO) and is dominated by the core diamagnetic contribution (the dotted line in Fig. S1) that is estimated as $-1.28 \times 10^{-4}$ emu/mol from the Pascal's constants [3]. From the Curie fitting of the data (the solid lines in Fig. S1), the Curie constant is estimated as $0.0032$ emu $\cdot$ K $\cdot$ (f. mol)$^{-1}$ and $0.0020$ emu $\cdot$ K $\cdot$ (f. mol)$^{-1}$ for $\chi$ measured under $B \parallel ab$ and $B \parallel c$, respectively. These estimations correspond to magnetic impurities about 0.3–0.4%, confirming the non-magnetic character of NZTO with negligible amount of the magnetic impurities. We note that it remains a future issue to investigate the dependence of the thermal Hall effect on the amount of magnetic impurities.

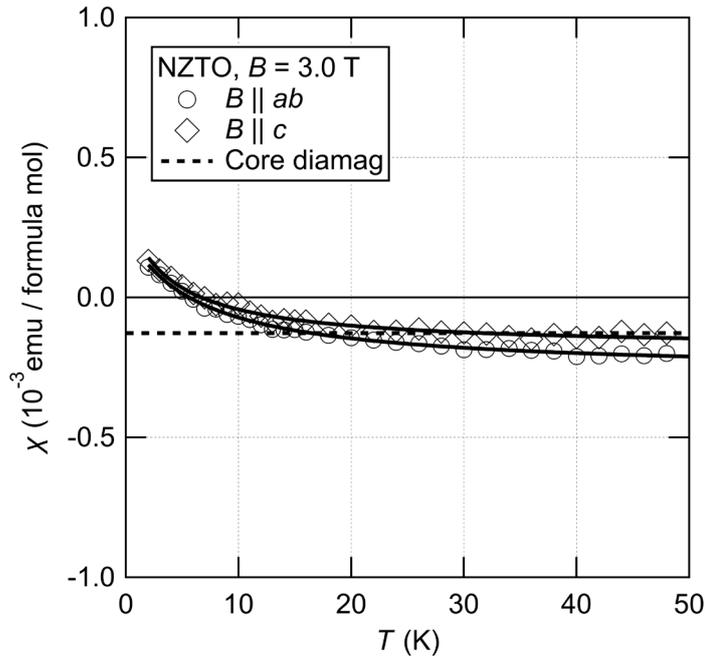

Fig. S1 The temperature dependence of the magnetic susceptibility of Na$_2$Zn$_2$TeO$_6$ measured at 3.0 T

for both $B \parallel ab$ (circles) and $B \parallel c$ (diamonds). The solid lines show Curie fits to the data.

## The field dependence of the magnetothermal conductivity in Na$_2$Co$_2$TeO$_6$ (NCTO) and Na$_2$Zn$_2$TeO$_6$ (NZTO)

As shown in Figs. 2(a–c) in the main text, the field dependence of the normalized magnetothermal conductivity $\Delta\kappa_{xx}(B)/\kappa_{xx}(0)$ of NZTO shows a positive magnetothermal conduction at lower temperatures and smaller angles, which is turned to be negative at higher temperatures and larger angles. This positive magnetothermal conduction is typical of the phonon thermal conduction that is enhanced by suppressing the scattering by magnetic impurities under magnetic field. The negative magnetothermal effect observed at higher temperatures would be due to a resonant scattering effect between the phonons and the Zeeman gap, which is known to block the phonon conduction [4].

In NCTO, $\Delta\kappa_{xx}(B)/\kappa_{xx}(0)$ shows the positive magnetothermal conductivity for the in-plane field, which turns into the negative one for the out-of-plane field, as shown in Figs. 3(a) and 3(b) in the main text. At higher temperatures, only the negative magnetothermal conductivity was observed for all the angles as shown in Fig. 3(c). As in the positive magnetothermal conduction in NZTO, this positive magnetothermal conduction is caused by an increase of the phonon contribution in $\kappa_{xx}$ by suppressing the spin fluctuation under the in-plane field [2,5]. Therefore, the decrease of the positive magnetothermal conduction by changing the field direction from the easy axis ($B \parallel a$) to the hard axis ($B \parallel c$) can be understood as a weakening of the suppression caused by the magnetic easy-plane anisotropy. The negative magnetothermal conduction observed in $B \parallel c$ is unlikely due to a phonon contribution because phonons show a positive magnetothermal conduction except for a resonance phonon effect [4]. This resonance phonon effect is reduced at higher temperature, which is opposite to the increase of negative magnetothermal conduction at 0° from 30 K to 46 K. Therefore, this negative magnetothermal conduction might be caused by a decrease of a magnetic contribution in $\kappa_{xx}$ as suggested by the previous work [6], in addition to the resonance scattering effect.

## The additional field dependence data of $\kappa_{xx}$ and $\kappa_{xy}$ of Na$_2$Co$_2$TeO$_6$ and Na$_2$Zn$_2$TeO$_6$ (Fig. S2 – S6).

Figure S2 shows the field dependence of the normalized magneto-thermal conductivity $\Delta\kappa_{xx}(B)/\kappa_{xx}(0) = \left(\kappa_{xx}(B) - \kappa_{xx}(0)\right)/\kappa_{xx}(0)$ and the thermal Hall conductivity divided by the temperature $\kappa_{xy}/T$ at 11 K. As shown in Fig. S2, the field dependence of both $\Delta\kappa_{xx}(B)/\kappa_{xx}(0)$ and $\kappa_{xy}/T$ shows a large magnetic hysteresis between the measurements done in the magnetization and the demagnetization processes. This large magnetic hysteresis at lower temperatures in the AFM phase makes it difficult to discuss the field dependence below 20 K.

Figure S3 shows the field dependence of $\Delta\kappa_{xx}(B)/\kappa_{xx}(0)$ at 22 K of this work (the red circles, from Fig. 3(a) in the main text) and at 20 K in the previous work [2], showing a good reproducibility.

Figures S4–S6 show additional data of the field dependence of $\kappa_{xy}$ used in Fig. 4 in the main text.

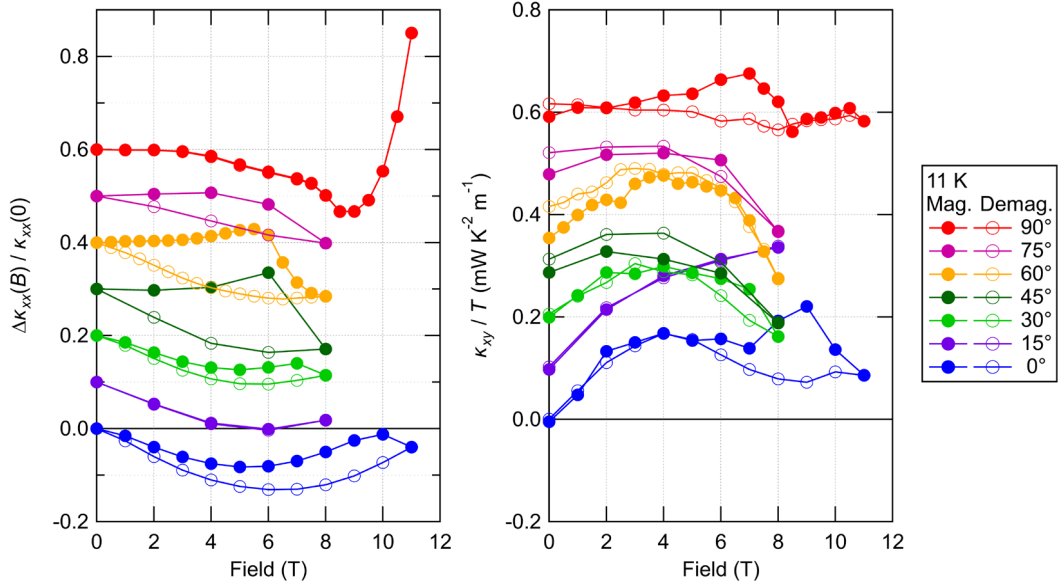

Fig. S2 The large magnetic hysteresis at low temperature in Na$_2$Co$_2$TeO$_6$. The magnetic field dependence of the normalized magneto-thermal conductivity $\Delta\kappa_{xx}(B)/\kappa_{xx}(0) = \bigl(\kappa_{xx}(B) - \kappa_{xx}(0)\bigr)/\kappa_{xx}(0)$ (left) and the thermal Hall conductivity divided by the temperature $\kappa_{xy}/T$ (right) at 11 K. The filled (open) symbols show the data obtained in the magnetization (demagnetization) process. The data in both panels is vertically shifted for clarity.

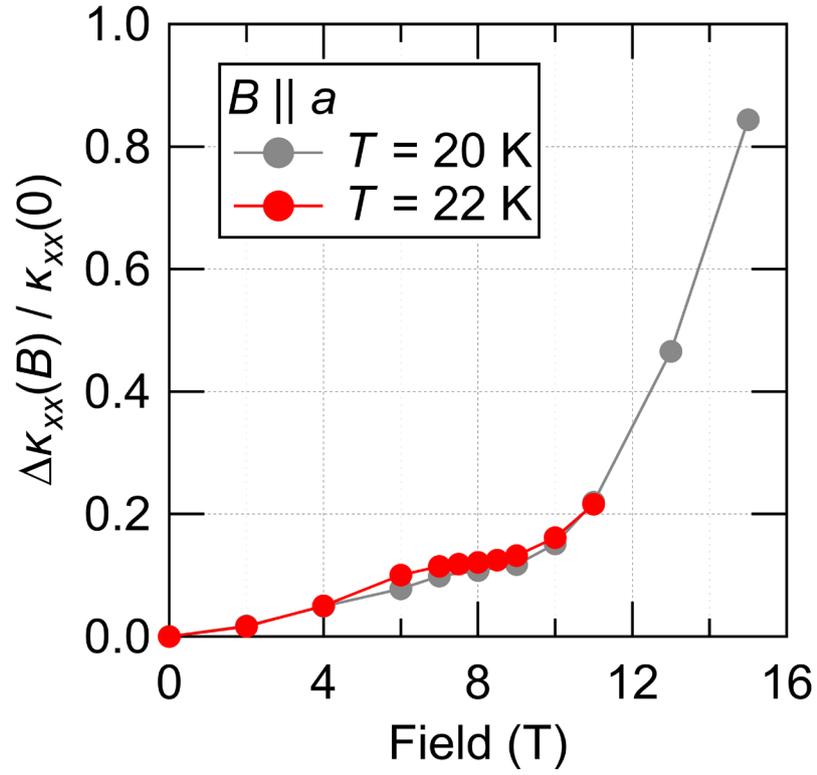

Fig. S3 Comparison of the magneto-thermal conductivity $\Delta\kappa_{xx}(B)/\kappa_{xx}(0) = (\kappa_{xx}(B) - \kappa_{xx}(0))/\kappa_{xx}(0)$ under the in-plane field of the current work (red circles, the data at 90° shown in Fig. 3(a)) and that of the previous work (grey circles from Ref. [2]).

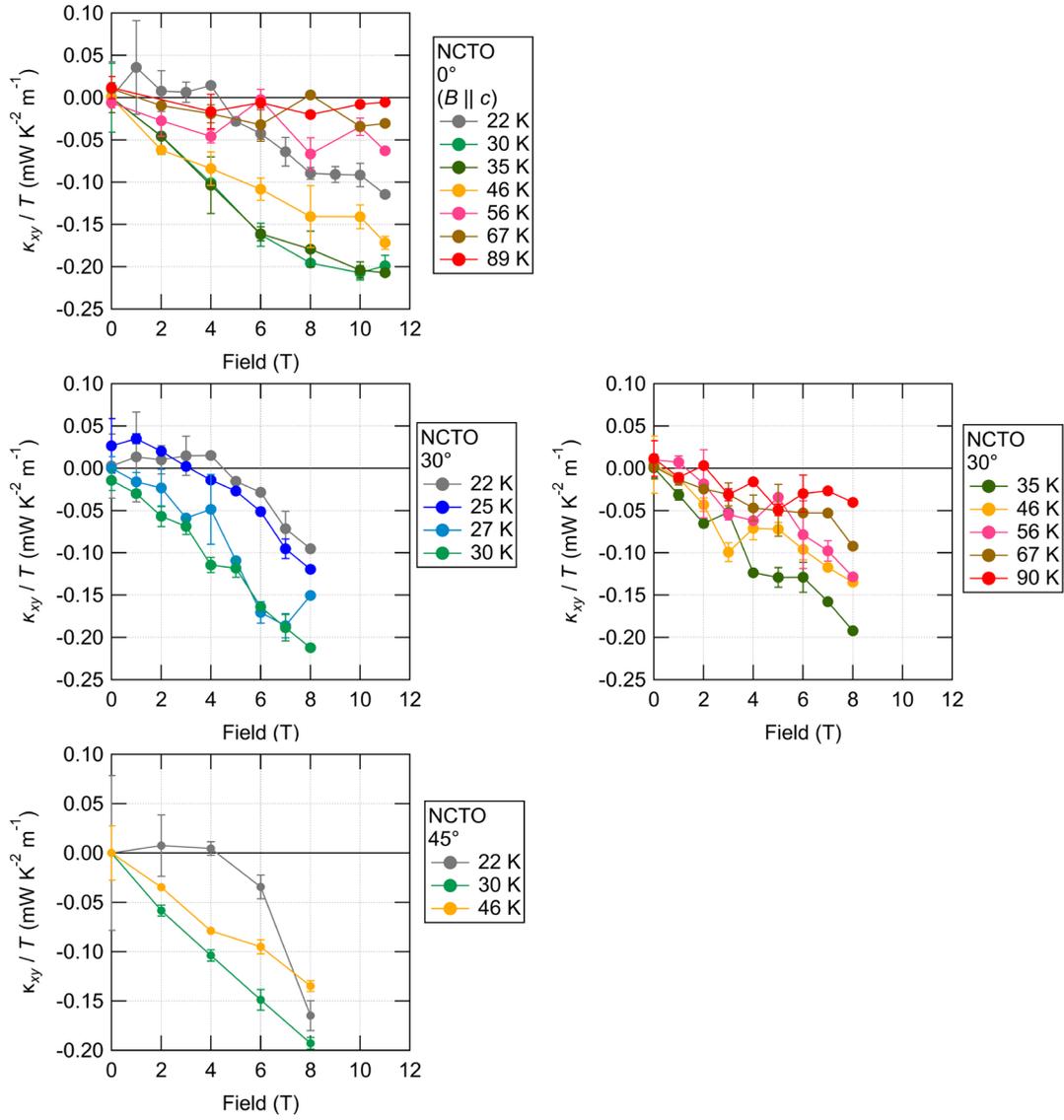

Fig. S4 The additional field dependence data of $\kappa_{xy}/T$ of Na$_2$Co$_2$TeO$_6$ at 0°, 30°, and 45°. The averaged data measured in the magnetization and the demagnetization processes is plotted. The error bars show the deviation of the data in the two processes.

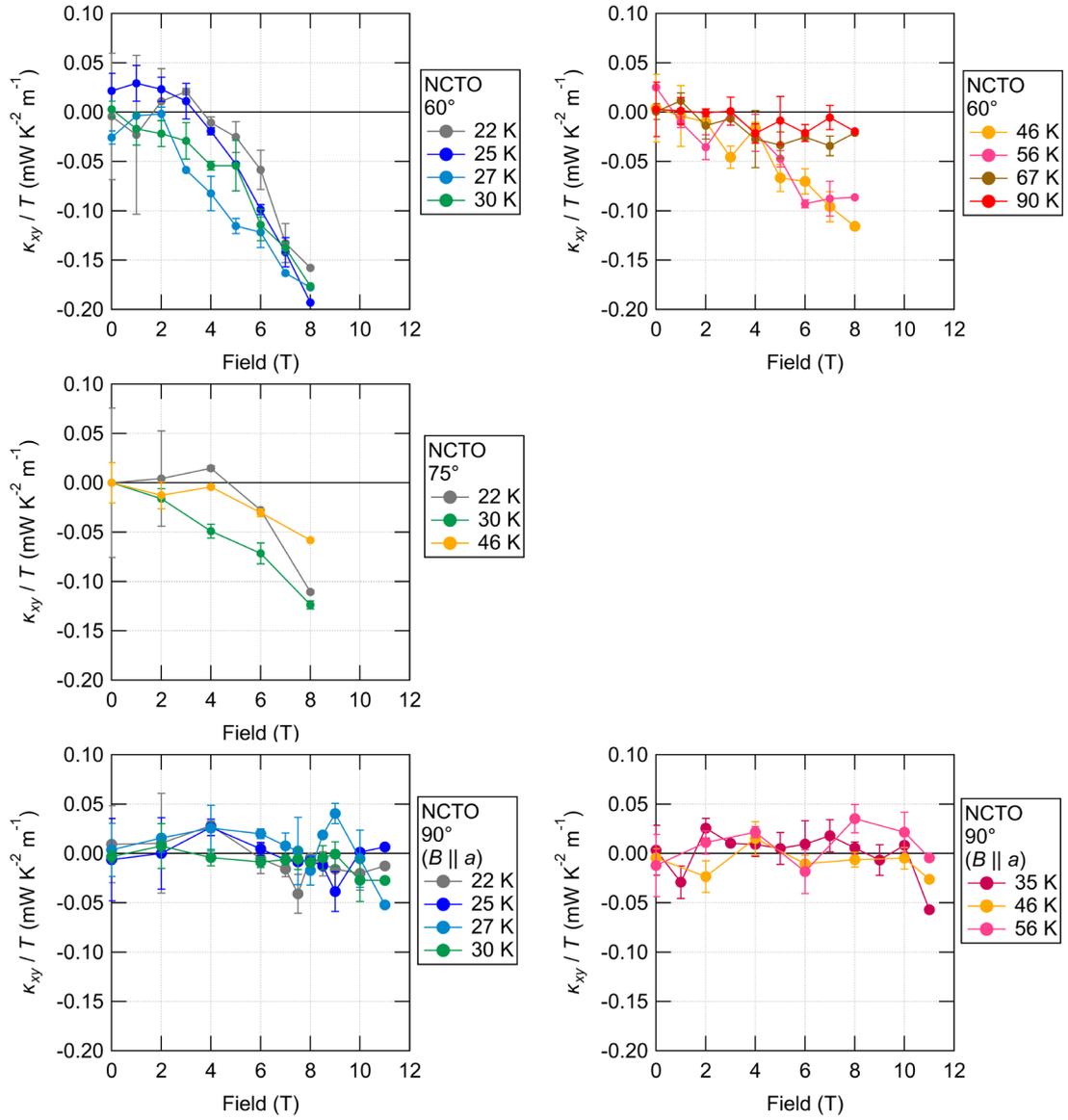

Fig. S5 The additional field dependence data of $\kappa_{xy}/T$ of $Na_2Co_2TeO_6$ at 60°, 75°, and 90°. The averaged data measured in the magnetization and the demagnetization processes is plotted. The error bars show the deviation of the data in the two processes.

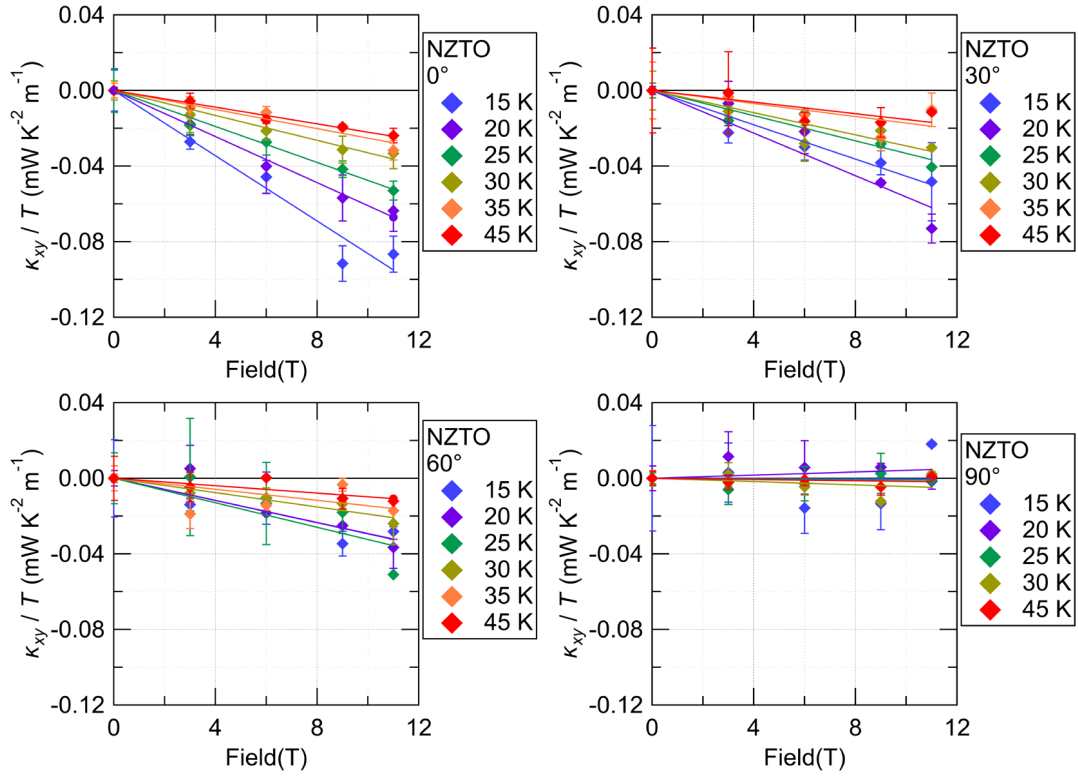

Fig. S6 The additional field dependence data of $\kappa_{xy}/T$ of $Na_2Zn_2TeO_6$. The averaged data measured in the magnetization and the demagnetization processes is plotted. The error bars show the deviation of the data in the two processes. The solid lines show a linear fit of the data.

**The additional field-angle dependence of $-\kappa_{xy}/TB$ and $\Delta\kappa_{xx}(B)/\kappa_{xx}(0)$ of $Na_2Co_2TeO_6$ and $Na_2Zn_2TeO_6$ (Fig. S7 and Fig. S8)**

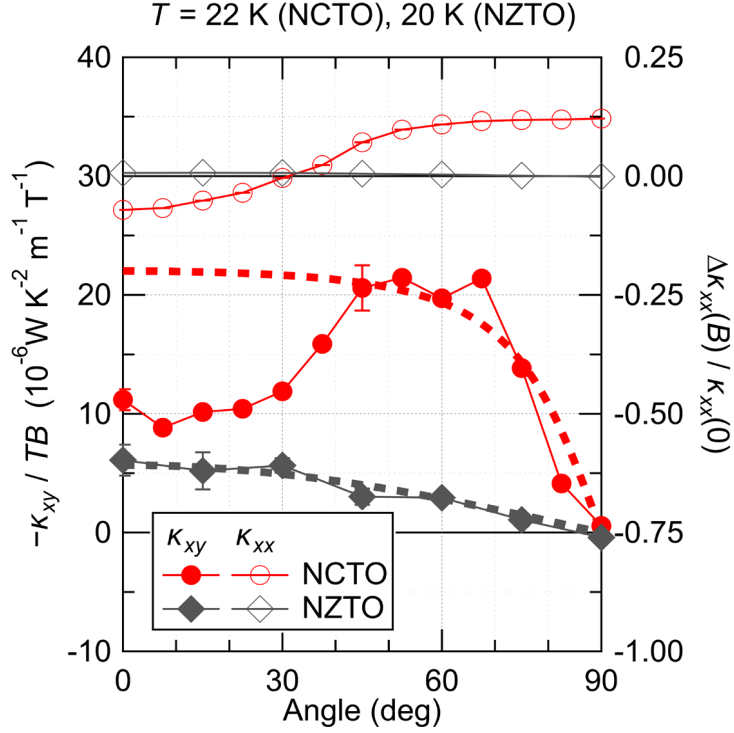

Fig. S7 The field-angle dependence of $-\kappa_{xy}/TB$ (filled symbols, left) and $\Delta\kappa_{xx}(B)/\kappa_{xx}(0)$ (open symbols, right) of NCTO (red) and NZTO (grey) at around 20 K. The $\Delta\kappa_{xx}(B)/\kappa_{xx}(0)$ data at 8 T (9 T) is shown for NCTO (NZTO). The definitions of the $-\kappa_{xy}/TB$ data and those of the error bars are the same as those used in Fig. 4 in the main text. The grey dashed line shows a cosine fit to the $-\kappa_{xy}/TB$ data of NZTO. The red dashed line shows a fit to the $-\kappa_{xy}/TB$ data of NCTO above 45° by the field-angle dependence of the out-of-plane magnetization $M_c(\theta) = M_c(0) \cdot (1 + (\tan\theta/\alpha)^2)^{-\frac{1}{2}}$, where $\alpha = 3.2$ is the averaged ratio of the magnetization along the $a$ axis to that along the $c$ axis of NCTO (Fig. 1(d) in the main text). Given that the transition from the antiferromagnetic state to the paramagnetic state increases $\kappa_{xx}$ [2,5], the decrease of $\Delta\kappa_{xx}(B)/\kappa_{xx}(0)$ below 45° shows the transition from the paramagnetic state at $\theta = 90°$ and at 8 T into the antiferromagnetic state [7]. Below this transition angle, the angle dependence of $|\kappa_{xy}/TB|$ shows a large suppression from the elliptic curve in the paramagnetic state ($\theta > 45°$). This suppression of $\kappa_{xy}$ in the antiferromagnetic state is opposite to that expected for $\kappa_{xy}$ from the extrinsic scatterings because the decrease of $\kappa_{xx}$ shows the increase of the phonon scatterings. Therefore, this suppression of $\kappa_{xy}$ suggests an emergence of another magnetic $\kappa_{xy} > 0$ in the antiferromagnetic state, which is consistent with the appearance of the negative $\Delta\kappa_{xx}(B)/\kappa_{xx}(0)$ for $\theta < 45°$.

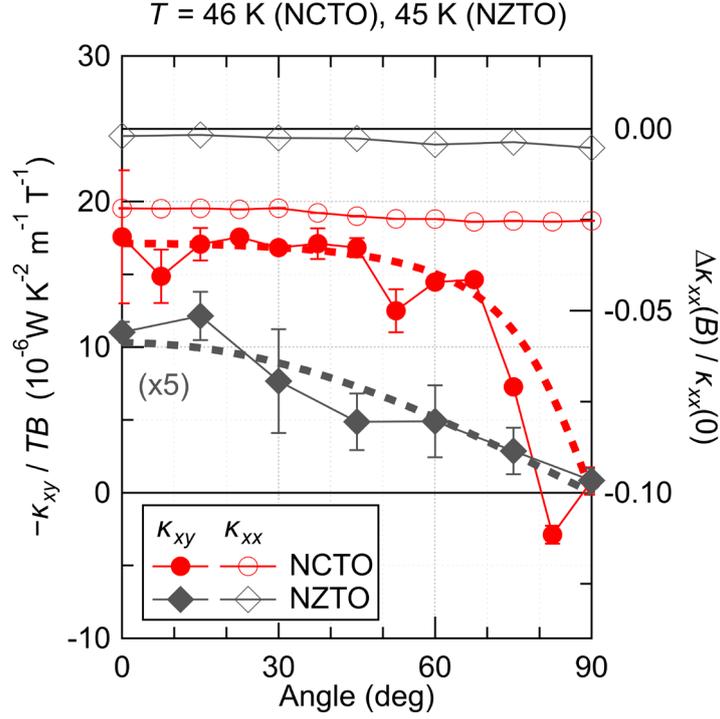

Fig. S8 The field-angle dependence of $-\kappa_{xy}/TB$ (filled symbols, left) and $\Delta\kappa_{xx}(B)/\kappa_{xx}(0)$ (open symbols, right) of NCTO (red) and NZTO (grey) at around 45 K. The $\Delta\kappa_{xx}(B)/\kappa_{xx}(0)$ data at 8 T (9 T) is shown for NCTO (NZTO). The definitions of the $-\kappa_{xy}/TB$ data and those of the error bars are the same as those used in Fig. 4 in the main text. The $-\kappa_{xy}/TB$ data of NZTO is multiplied by 5 for clarity. The grey dashed line shows a cosine fit to the $-\kappa_{xy}/TB$ data of NZTO. The red dashed line shows a fit to the $-\kappa_{xy}/TB$ data of NCTO above 45° by the field-angle dependence of the out-of-plane magnetization $M_c(\theta) = M_c(0) \cdot (1 + (\tan\theta/\alpha)^2)^{-\frac{1}{2}}$.